\documentclass[prc,aps,showpacs,floatfix,nofootinbib,twocolumn,letterpaper]{revtex4}
\usepackage{graphicx}
\usepackage{amssymb}
\usepackage{amsmath}
\usepackage{amsfonts} 
\usepackage{array} 
\usepackage{tabularx}
\newcolumntype{R}{>{\raggedleft\arraybackslash}X}
\newcolumntype{L}{>{\raggedright\arraybackslash}X}
\newcolumntype{C}{>{\centering\arraybackslash}X}

\newcommand{\la}{\langle}
\newcommand{\ra}{\rangle}
\newcommand{\be}{\begin{equation}}
\newcommand{\ee}{\end{equation}}
\newcommand{\ba}{\begin{eqnarray}}
\newcommand{\ea}{\end{eqnarray}}
\newcommand{\ad}{a^\dagger}

\newcommand{\ve}{\varepsilon}
\newcommand{\gs}{|\Psi_{\rm gs} \rangle}

\newcommand{\dt}{\delta \tau}
\newcommand{\Dt}{\Delta \tau}
\newcommand{\bn}{{\bf n}}
\newcommand{\bm}{{\bf m}}
\newcommand{\mh}{\mathcal{H}}

\newcommand{\mS}{\mathcal{S}}
\newcommand{\mE}{\mathcal{E}}
\newcommand{\mP}{\mathcal{P}}
\newcommand{\ua}{\uparrow}
\newcommand{\da}{\downarrow}
\newcommand{\bi}{\bibitem}

\begin{document}
\title{Configuration-interaction Monte Carlo method and its
application to the trapped unitary Fermi gas}
\author{Abhishek Mukherjee$^{1,2}$ and Y. Alhassid$^1$}
\affiliation{$^1$Center for Theoretical Physics, Sloane Physics
Laboratory, Yale University, New Haven, CT 06520, USA\\
$^2$ECT*, Villa Tambosi, I-38123 Villazzano, Trento, Italy}
\begin{abstract}
\noindent We develop a quantum Monte Carlo method to estimate the ground-state energy of a fermionic many-particle system in the configuration-interaction shell model approach. The fermionic sign problem is circumvented by using a guiding wave function in Fock space. The method provides an upper bound on the ground-state energy whose tightness depends on the choice of the guiding wave function. We argue that the antisymmetric geminal product class of wave functions is a good choice for guiding wave functions. We demonstrate our method for the trapped two-species fermionic cold atom system in the unitary regime of infinite scattering length using
the particle-number projected Hartree-Fock-Bogoliubov wave function as the guiding wave function.
 We estimate the ground-state energy and energy-staggering pairing gap as a function of the number of particles. Our
  results compare favorably with exact numerical diagonalization results and with
  previous coordinate-space Monte Carlo calculations.
 \end{abstract}

\maketitle
The configuration-interaction (CI) shell model
approach is widely used in nuclear, atomic and molecular physics. It accounts for
both shell effects and correlations in finite-size many-particle systems (see,
e.g., Refs.~\onlinecite{ism,ci}). The CI many-particle model space for an
$N$-fermion system is a truncated space spanned by $N$-particle Slater determinants
 constructed from a given finite single-particle basis.
 An effective CI Hamiltonian is defined in this truncated space.

When the many-particle model space is sufficiently small, the CI Hamiltonian can be diagonalized by conventional methods. However, the combinatorial increase of the dimensionality of the many-particle
space versus the size of the single-particle basis and/or the number
of particles prohibits direct diagonalization in large spaces.
This difficulty can be overcome in part by using quantum Monte Carlo methods for
which the computational effort scales much more gently with the size of the
single-particle model space. An example is the auxiliary-field Monte Carlo (AFMC) method. The AFMC approach
has been applied within the CI framework to nuclei~\cite{la93,al94,koo97,al01}
(where the method is known as the shell model Monte Carlo method), and more
recently to the cold atomic condensate in a harmonic trap~\cite{gil12b}.

Other quantum Monte Carlo methods include coordinate-space (`$\mathbf{r}$-space') Monte Carlo methods, e.g., diffusion Monte Carlo and Green's function Monte Carlo (GFMC)~\cite{qmc99}. These methods filter out the ground state with the help of an appropriately defined projection operator or a propagator. A GFMC method that works on a discrete lattice  in coordinate space (known as lattice GFMC) was also developed~\cite{bem94,ten95,sor00}.

Here we introduce a novel configuration-interaction Monte Carlo (CIMC) method that
 provides an upper bound of the ground-state energy of a fermionic system in
the CI framework.
Our method builds on techniques developed in the lattice GFMC method.
 A discrete configuration space is defined by the occupation numbers of the single-particle states and an initial configuration-space wave function is propagated in imaginary time. The sign problem is circumvented by introducing a propagator that depends on a guiding wave function and keeps the wave function positive semidefinite. The method yields an upper bound on the ground-state energy whose accuracy (as an estimate of the ground-state energy) depends on the choice of the guiding wave function. We argue that the use of antisymmetric geminal product (AGP) wave functions~\cite{col00} as guiding wave functions offer a good compromise between accuracy and computational efficiency.

We demonstrate the CIMC method by its application to the trapped two-species
fermionic cold atom system in the unitary limit of infinite scattering length. We use the particle-number projected Hartree-Fock-Bogoliubov (HFB) wave function
(a member of the AGP class) as the parameter-free guiding wave function.
Cold atomic gases have recently attracted much interest both experimentally and
theoretically~\cite{Bloch2008,gio08}. The interaction strength in these systems
can be controlled by tuning the scattering length near a Fano-Feshbach
resonance. Since these systems depend on a few number of parameters, they are useful for testing many-body methods of strongly interacting systems. The unitary limit is particularly challenging since, in the
absence of a small parameter, it is not amenable to perturbative treatments.

Our method consists of two main components: (i) projecting on the ground-state wave function
through a random walk in configuration space, and (ii) circumventing the fermionic sign problem with the help of a guiding wave function.

 We assume a general CI Hamiltonian which includes only two-body interactions
\be
H=\sum_{a \in \mS} \ve_a \ad_a a_a + \sum_{abcd \in \mS} V_{abcd}\ad_a \ad_b a_c a_d \;,
\label{eqnham1}
\ee
where $\ad_a$ creates a particle in the single-particle
state labeled by $a$.  The set $\mS$ of single-particle states is assumed to be
finite of size ${\cal N}_s$.

We define an operator $\mP$  by
\be
\mP = 1 - \Dt (H-E_T) \;,
\ee
where $E_T$ is an energy shift (to be discussed later). This operator can be used to propagate in imaginary time a wave function in the many-particle space from $\tau$ to $\tau+\Dt$  by
\be
\left | \Psi_{\tau + \Dt} \right \ra = \mP \left |\Psi_\tau \right \ra\;.
\label{pro1}
\ee
The ground-state wave function $\gs$ can be projected out by the repeated
application of $\mP$ on an initial wave function $\Psi_0$ that has a non-zero overlap with $\Psi_{\rm gs}$, i.e.,
\be \gs = \lim_{\tau \to \infty} |\Psi_\tau \ra \;,
\label{pro2}
\ee
provided the eigenvalues of $\mP$ are between $-1$ and $1$. The latter condition implies that the time step $\Dt$ satisfies $\Dt < 2/(E_{\rm max} - E_T)$, where $E_{\rm max}$ is the maximal eigenvalue of $H$. Equation (\ref{pro2}) is exact and there is no error that depends on the size of the time step.

 The method works in the $N$-particle Hilbert space that is spanned by the set of all $N$-particle Slater determinants constructed from the single-particle orbitals $a \in \mS$. We will denote these Slater determinants or `configurations' by $|\bn \ra$, where $\bn \equiv \{ n_a\}$ and $n_a = 0,1$ is the occupation number of 
orbital $a$. The one-step propagation (\ref{pro1}) can be written in the configuration representation as
\be
\Psi_{\tau + \Dt}(\bm) = \sum_{\bn} \la \bm |\mP | \bn \ra  \Psi_\tau(\bn)\;;
\label{config-pro}
\ee
where $\Psi_\tau(\bn) \equiv \la \bn | \Psi_\tau \ra$ is the wave function representation in configuration space.
We rewrite the configuration-space matrix elements of $\mP$ in the form
\be
\la \bm |\mP | \bn \ra = g(\bn)p(\bm,\bn) \;,
\label{eqnpr1}
\ee
where
\be
 g(\bn)     = \sum_{\bm} \la \bm |\mP | \bn \ra  \;,
\label{eqnpr2}
\ee
and
\be
 p(\bm,\bn) = \frac{\la \bm |\mP | \bn \ra}{g(\bn)} \;.
\label{eqnpr3}
\ee

We first consider the case when the matrix elements of $\mP$ are all non-negative, i.e.,  $\langle \bm|\mP|\bn\rangle  \geq 0$ for all $\bm$ and $\bn$. Then $0\leq p(\bm,\bn)\leq 1$ with $\sum_{\bm}p(\bm,\bn)=1$ and $g(\bn)\geq 0$, and we can interpret $p(\bm,\bn)$ for fixed $\bn$ as a probability and $g(\bn)$ as a weight. This enables us to carry out the propagation of $\Psi_\tau$ in (\ref{config-pro}) stochastically as follows. Assuming at a given time $\tau$, the wave function $\Psi_\tau$ is non-negative in configuration space, i.e., $\Psi_\tau(\bn) \geq 0$ for all $\bn$, the wave function $\Psi_\tau$ can be represented as an ensemble of configurations $\bn$. According to (\ref{config-pro}) and the non-negativity of the matrix elements of $\mP$, the wave function remains non-negative  at $\tau+\Dt$, i.e., $\Psi_{\tau+\Dt}(\bm)\geq 0$ for all $\bm$.
For each configuration $\bn$, a new configuration $\bm$ is chosen with probability
$p(\bm,\bn)$ and replicated with weight $g(\bn)$. The resulting ensemble of configurations $\{\bm\}$ samples the wave function  $\Psi_{\tau + \Dt}$ at the next imaginary-time step $\tau+\Dt$. We note that in CIMC, the diffusion in configuration space is determined by the interaction part of the Hamiltonian, while in coordinate-space Monte Carlo methods it is the kinetic part which governs the diffusion.

After a sufficiently large number of time steps, the contribution from excited
 states to $\Psi_\tau$ becomes negligible. Ensembles generated at subsequent time
steps are considered as representatives of $\gs$ and a decorrelated subset of
these ensembles is used to calculate the observables.

As mentioned above, the propagation with $\mP$ filters out the ground state when the time step satisfies $\Dt < 2/(E_{\rm max} - E_T)$. This upper bound for $\Dt$ becomes smaller with increasing $N$ and/or ${\cal N}_s$ since $E_{\rm max} $ gets larger. Consequently, the number of time steps required for decorrelation and for projecting the ground state increases. This makes the simple algorithm described above inefficient for large $\mathcal{N}_s$ and/or $N$.

We overcome this latter difficulty by using an algorithm proposed in Ref.~\onlinecite{tri90}.
We start by choosing a finite imaginary time step $\dt$ (which is different from $\Dt$). Assuming the configuration $\bn$ is a member of the ensemble
at time $\tau$, we describe the choice of the corresponding configuration and its replication weight at time $\tau + \dt$.  We define a propagation time $\dt_p$ that is initially set to $\dt$ and a weight $g$ that is initially set to 1. We then sample a time $\dt_d$ for an off-diagonal move from the (unnormalized)
probability distribution $e^{-\pi_d \dt_d}$, where $\pi_d=\sum_{\bm \neq \bn}
\langle \bm | H-E_T|\bn \rangle$.  If $\dt_d \ge \dt_p$, then $\bn$ remains unchanged in the next time step and is replicated with the weight $g = e^{-\dt_p \sum_{\bm} \langle \bm | H - E_T | \bn \rangle}$.
Otherwise, a new configuration $\bn'$ is chosen with probability $\langle \bn'| H -E_T|\bn \rangle /\pi_d$ $(\bn' \neq \bn)$,
and the weight factor is multiplied by $e^{-\dt_d\sum_{\bm} \langle \bm | H - E_T| \bn \rangle}$. This process is repeated with the replacements
  $\bn \to \bn'$ and $\dt_p \to \dt_p - \dt_d$ until we have $\dt_d \ge \dt_p$.
  This algorithm generates the probability distribution $p(\bm , \bn)$  and
  weight $g(\bn)$ that correspond to the propagator $e^{-\dt (H - E_T)} = \lim_{\Dt \to 0}\mP^{\dt/\Dt}$
 without any time step errors due to the finiteness of $\dt$~\cite{sor00}.

The above algorithm breaks down if $p(\bm,\bn) < 0$, i.e., if
$\langle \bm |H | \bn \rangle > 0$ for any pair of configurations $\bm \neq \bn$. For a general CI Hamiltonian, this might be the case and is the manifestation of the Monte Carlo sign problem for our method~\footnote{An important exception is the pairing Hamiltonian, for which a diffusion Monte Carlo algorithm free of a sign problem can be formulated within the CI framework using the occupations of time-reversed pairs~\cite{cer93,muk11}.}

In continuum $\mathbf{r}$-space Monte Carlo methods the sign problem is
circumvented with the help of a fixed-node approximation, which can be used
to obtain an upper bound on the ground-state energy. However, in the CI approach 
the Hilbert space is labeled by discrete quantum numbers. It was shown in lattice GFMC that a fixed-node-like approximation can be introduced
to obtain an upper bound on the ground-state energy also for discrete Hilbert spaces.
In the following we show how this can be formulated for a CI Hamiltonian.

We choose a guiding wave function $\Phi_G$ and define for any configurations $\bn$ and $\bm$ with $\Phi_G(\bn) \neq 0$ the quantity 
$\mathfrak{s}(\bm,\bn) \equiv \Phi_G(\bm) \langle \bm| H |\bn \rangle /\Phi_G(\bn)$.
We then introduce a family of Hamiltonians $\mh_{\gamma}$ defined over configurations $\bn$  with $\Phi_G(\bn) \neq 0$ such that the off-diagonal matrix
elements are given by~\cite{bem94,ten95,sor00}
\begin{align}
  \label{mh1}
  \langle \bm |\mh_{\gamma} |\bn \rangle  =\left \{ \begin{array}{rr} -\gamma \langle \bm| H |\bn \rangle  & \mathfrak{s}(\bm,\bn)  > 0 \\
  \langle \bm| H |\bn \rangle   &  \mbox{otherwise} \end{array} \right . \;,
\end{align}
while the diagonal matrix elements are
\be
\label{mh2}
\langle \bn | \mh_{\gamma} | \bn \rangle = \langle \bn | H | \bn \rangle+ (1+\gamma) \displaystyle \sum_{\stackrel{ \bm \neq
\bn}{\mathfrak{s}(\bm,\bn) > 0}} \mathfrak{s}(\bm,\bn) \;.
\ee
We note that $\mh_{\gamma = -1} =H$. We define a $\gamma$-dependent propagator $\mP_{\gamma}$ for configurations $\bn$ with $\Phi_G(\bn) \neq 0$ by
\be
\label{mpg}
\langle \bm | \mP_{\gamma} |\bn \rangle = 1 - \Dt \Phi_G (\bm) \langle \bm | \mh_{\gamma} - E_T|\bn \rangle /\Phi_G(\bn)\;.
\ee
For $\gamma \geq 0$  the propagator $\mP_{\gamma}$ satisfies $\langle \bm | \mP_{\gamma}|\bn \rangle \geq 0$ and is therefore free of the sign problem. The stochastic projection algorithm outlined above can then be generalized with the replacement of $\langle \bm| H |\bn \rangle$ by $\Phi_G (\bm) \langle \bm | \mh_{\gamma}|\bn \rangle /\Phi_G(\bn)$.
This stochastic projection filters out the wave function $\Phi_G(\bn) \Psi_{\gamma}(\bn)$, where
$\Psi_{\gamma}(\bn)$ is the ground-state wave function of $\mh_{\gamma}$.

The ground-state energy $\mathcal{E}_{\gamma}$ of $H_{\gamma}$ (in the non-singular space $\Phi_G(\bn) \neq 0$) is
an upper bound for the ground-state energy $E_0$ of the original Hamiltonian $H$ for
$\gamma \geq 0$. This result is derived in Refs.~\onlinecite{ten95} and \onlinecite{sor00}
for the case when $\Phi_G$ is nonzero for all configurations in the Hilbert space.
In the following we derive this result for the case where $\Phi_G(\bn)$ may vanish
 for certain configurations (this is typically the situation for CI guiding wave functions).
 In this case the propagator $\mP_{\gamma}$
 is non-singular only in the subspace of configurations $\bn$ spanned by $\Phi_G(\bn) \neq 0$.
The Hamiltonian $\mh_{\gamma}$ is also non-singular in this subspace.
 The energy $\mathcal{E}_{\gamma}$ is an upper bound for the ground-state energy of $H$ when
restricted to this non-singular subspace~\cite{ten95,sor00}. This latter energy is an upper
bound on $E_0$ (which is the ground-state energy of $H$ in the full Hilbert space). Thus, $\mathcal{E}_{\gamma}$ is an upper bound on $E_0$.

It is seen from Eq.~(\ref{mpg}) that as long as the initial ensemble includes only those
configurations for which $\Phi_G(\bn) \neq 0$, the Monte Carlo projection does not reach
configurations $\bm$ for which $\Phi_G(\bm) = 0$. Thus, the Monte Carlo projection correctly
 finds the ground-state energy in the space where $\mP_{\gamma}$ is non-singular and thus provides an upper bound on $E_0$.

One can verify using Eqs.~(\ref{mh1}) and (\ref{mh2}) that  $\langle \Phi_G | \mh_{\gamma} | \Phi_G \rangle = \langle \Phi_G| H |\Phi_G \rangle $.
 Since, $\mathcal{E}_{\gamma}$ is the ground-state energy of $\mh_{\gamma}$,
$\mathcal{E}_{\gamma} \leq \langle \Phi_G|\mh_{\gamma}|\Phi_G \rangle =\langle \Phi_G|H|\Phi_G \rangle $,
i.e., the upper bounds on $E_0$ given by $\mathcal{E}_{\gamma}$ are tighter than the variational upper bound with $\Phi_G$.

The energies $\mathcal{E}_\gamma$ are estimated using the `mixed' estimate for the Hamiltonian $\mh_\gamma$
\be
\mE_{\gamma} =  \frac{\displaystyle \sum_{\bn} \mE^L(\bn) \Phi_G(\bn)\Psi_{\gamma}(\bn)}{\displaystyle \sum_{\bn}
\Phi_G(\bn)\Psi_{\gamma}(\bn)} \approx \frac{1}{N_E} \sum_i \mE^L(\bn_i)
\label{eqnen},
\ee
where $N_E$ is the size of the ensemble $\{\bn_i\}$ representing
$\Phi_G(\bn)\Psi_{\gamma}(\bn)$ and $\mE^L(\bn) = \langle \Phi_G |\mh_{
\gamma} |\bn \rangle /\Phi_G(\bn) = \langle \Phi_G |H|\bn \rangle/\Phi_G(\bn) $
is the so-called local estimate of the energy.

The energy shift $E_T$ is used to control the size of the configuration population. It
is adjusted infrequently during the evolution to keep the population of configurations
roughly constant. It provides an independent estimate (the  `growth'
estimate) of the ground-state energy of $\mh_\gamma$. However, its statistical error is typically much larger than that of the `mixed' estimate described above.

The finite time step $\dt$ is kept fixed throughout the entire calculation.
In principle, its value is arbitrary since the method is free
from time step errors. However, if $\dt$ is too large, then the configuration
population undergoes large fluctuations between consecutive time steps.
On the other hand, if $\dt$ is too small, then the number of time steps
required to decorrelate the energy becomes very large. As a compromise we choose an intermediate value for $\dt$.

A linear extrapolation of $\mE_{\gamma}$ from any two values of $\gamma \geq 0$ to $\gamma = -1$ also provides
an upper bound on $E_0$ that is tighter than the individual $\mE_{\gamma}$'s~\cite{bec01}.
We found that the best compromise between the tightness of the upper bound and the size of the statistical error in
the extrapolation is obtained when the two values of $\gamma$ are chosen to be $0$ and $1$. This choice gives
$E_{\rm CIMC} = 2\mE_{\gamma = 0} - \mE_{\gamma = 1}$ as our best upper bound for the ground-state energy $E_0$.

The tightness of the energy upper bound is determined to a large extent by the quality
of the guiding wave function.
If the guiding wave function is the exact ground-state wave function
(i.e., $ \Phi_G = \Psi_{\rm gs}$), then $\mE_{\gamma}= E_0$ for
all $\gamma$. Furthermore, for $\Phi_G$ close to $\Psi_{\rm gs}$ the difference
$\mE_{\gamma} - E_0$ is second order in $\Phi_G - \Psi_{\rm gs}$~\cite{ten95}. Thus,
the accuracy of the CIMC method can be improved by systematically improving the
quality of the guiding wave function.

Ideally, $\Phi_G$ should encode all our \emph{apriori} knowledge about the ground-state wave function. However, for the
CIMC method to be efficient, we should be able to calculate $\Phi_G$ quickly and accurately.
As we discussed above, the configuration space in which the stochastic projection is carried
 out is determined by the condition $\Phi_G \neq 0$. It is preferable for this space to be
sufficiently large so as to include the dominant correlations.

In $\mathbf{r}$-space Monte Carlo methods, optimized Slater-Jastrow or BCS-Jastrow wave functions are examples of highly accurate yet efficient guiding wave functions. For a CI Hamiltonian, a good choice
 is provided by the AGP class of wave functions~\cite{col00}. The most general AGP wave function
 for an even $N$-particle system is given by
\be
|\Phi_{\rm AGP}\rangle = \left (\sum_{ab} \phi_{ab} \ad_a \ad_b \right )^{N/2} |0\rangle \;,
\label{agp}
\ee
where $|0\rangle$ is the particle vacuum. The coefficients $\phi_{ab}$  are the `geminals' which encode information about the correlations in the system. 
 The AGP wave functions incorporate
important two-body correlations yet they are described by a single pfaffian, i.e.,
$\Phi_{\rm AGP}(\bn)$ is the pfaffian of an $N \times N$ matrix \cite{pfaff}. This later property ensures
their efficient numerical evaluation. More recently, AGP wave functions have also been used
extensively in $\mathbf{r}$-space quantum Monte Carlo calculations (see, e.g., in Ref.~\onlinecite{kol11}).

To demonstrate the CIMC method, we consider the two-species (labeled by spin up and spin down) fermionic cold atom system, in which atoms with opposite spins interact via a short-range interaction, modeled by a contact interaction
$\delta({\bf r - r'})$. The CI many-body Hamiltonian of this system in an isotropic harmonic trap is given by
\begin{widetext}
\be
H=\sum_{i \in \mS} \ve_i \left ( \ad_{i\ua} a_{i\da} + \ad_{i\da} a_{i\da} \right ) + g  \hbar \omega \left( \frac{\hbar}{m \omega }\right)^{3/2} \sum_{ijkl \in \mS} \la ij|\delta({\bf r - r'})|kl\ra\ad_{i\ua}\ad_{j\da} a_{l\da} a_{k\ua};,
\ee
\end{widetext}
where $\omega$, $m$ and $\ve_i$ are, respectively, the frequency of the trap, the particle mass and the single-particle energies in an isotropic harmonic trap.  The label $i$ denotes a single-particle state in orbital space and $\ua,\da$ denote spin-up and spin-down particles.

The dimensionless coupling strength $g$ for finite ${\cal N}_s$ is determined by
the condition that the two-particle ground-state energy in the laboratory frame
reproduces the exact energy~\cite{gil12b}. In the unitary limit of infinite
scattering length the exact two-particle ground-state energy is
$2\hbar\omega$~\cite{bus98}. We note that the renormalization of the contact
interaction in a finite model space is a non-trivial problem and there exist more rigorous
treatments of the effective interaction~\cite{ste07,alh08,gil12a}.

We choose the single-particle model space $\mS$ to include all single-particle
orbitals within ${\cal N}_{\rm max}$ harmonic oscillator shells, i.e, with
energy $ \ve_i \leq ({\cal N}_{\rm max}+\frac{3}{2}) \hbar \omega$. A model
space with ${\cal N}_{\rm max} = 9$ has ${\cal N}_s=440$ single-particle states.
The respective many-particle space for $N=20$ particles has dimension
of $\sim 10^{35}$.

For this system it is reasonable to assume that the dominant correlations
are in the pairing and particle-hole channels. A wave function that includes
these correlations is described by the HFB approximation.
The HFB wave function does not conserve particle number, thus it is
necessary to project it on a fixed number of particles.
We only consider here even-$N$ spin-balanced systems and odd-$N$ systems with
one unbalanced spin up particle.
The particle-number projected Hartree-Fock-Bogoliubov (PHFB) wave function (projection after variation) belongs to the AGP class of wave functions.
For the even-$N$ ($N=2n$) system it is given by
\be
\label{phfbe}
|\Phi^e_{\rm PHFB} \ra = \left (
\sum_{ijk} \frac{v_k}{u_k} D_{ik}D_{jk} \ad_{i\ua} \ad_{j\da} \right
)^n |0 \ra \;,
\ee
where the matrix $D$ describes
the transformation to the canonical basis, and $(u_k,v_k)$ are the coefficients
of the Bogoliubov transformation in the canonical basis~\cite{rin80}.

For odd-$N$ ($N=2n+1$) system, the wave function belongs to the generalized AGP
 class of wave functions and is given by
\be \label{phfbo}
|\Phi^o_{\rm PHFB} \ra = \left ( \sum_i D_{ib} \ad_{i\ua} \right ) \left (
\sum_{ij,k \neq b} \frac{v_k}{u_k} D_{ik}D_{jk} \ad_{i\ua} \ad_{j\da} \right
)^n |0 \ra \;,
\ee
where $b$ is the `blocked' spin-up orbital in the canonical basis. The configuration-space representations of
$\Phi^e_{\rm PHFB} (\bn)$ and $\Phi^o_{\rm PHFB} (\bn)$  are
determinants of $n \times n$ and $(n+1) \times (n+1)$ matrices, respectively.

The particle-number projected Bardeen-Cooper-Schrieffer (PBCS) and Hartree-Fock (HF)
wave functions have similar forms. In PBCS the $D$ matrix is the identity, while in
HF $v_k/u_k = \theta (k_F-|k|)$, with $k_F$ being the highest occupied orbital in the
canonical basis. Generally, the quantities $D_{ij}$, $u_k$ and $v_k$ have to be recalculated self-consistently
depending on which wave function we use, e.g., the $u_k,v_k$ in PHFB are not the same as those in PBCS.

\begin{figure}[tb]
\includegraphics[width=\columnwidth]{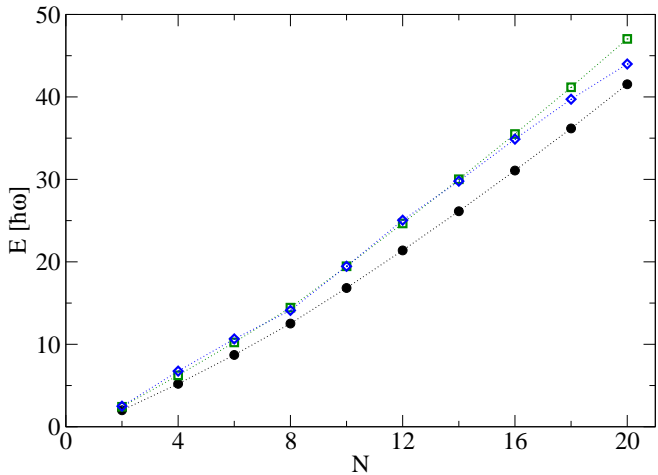}
\caption{The CIMC
    ground-state energy estimates versus particle number $N$ for ${\cal N}_{\rm max} = 6$ at unitarity using the
    renormalized Hamiltonian described in the text for different choices of the
    guiding wave function: PHFB (solid circles), PBCS (open squares), and HF (open diamonds). The statistical errors are smaller than the size of the symbols in all cases.}
  \label{figbcshf}
\end{figure}
The PBCS wave function includes correlations in the pairing channel only, while the HF wave function includes correlations in the particle-hole channel only. We expect the PHFB wave function, which includes correlations in both channels to be superior to either of those. This expectation is confirmed in Fig.~\ref{figbcshf},
where we compare the CIMC ground-state energy estimates at the unitary limit for different choices of the guiding wave function: HF, PBCS and PHFB in a truncated model space of ${\cal N}_{\rm max} = 6$ oscillator shells.
The energy estimates provided by the PHFB guiding wave function are much lower than
those of the HF and PBCS wave functions for all values of the particle number $N$.

\begin{table}[htbp] \begin{tabularx}{0.8\columnwidth}{CCRR}
  \hline\noalign{\smallskip} &   & \multicolumn{2}{c}{Ground state energy
  $[\hbar \omega]$} \\ \cline{3-4}\noalign{\smallskip} ${\cal N}_{\rm max}$ & N
  & CIMC & Exact \\ \hline\noalign{\smallskip} 3          & 6  &  8.639(7) &
  8.601      \\ & 7  & 11.176(4) & 11.021      \\ & 8  & 12.292(7) & 12.179
  \\ 8          & 2  &  2.000(6) &  2.000      \\ & 3  &  4.391(4) &  4.279
  \\ & 4  &  5.208(9) &  5.138      \\ \hline \end{tabularx}
\caption{Comparison
    of the CIMC ground-state energies versus the exact ground-state energies for a CI
    model space of ${\cal N}_{\rm max}=3$ and ${\cal N}_{\rm max}=8$, and for several
    values of the particle number $N$. For the CIMC energies, the numbers in
    parenthesis denote the statistical error in the last significant digit.  The
    ${\cal N}_{\rm max}=3$ energies were obtained using the {\tt oxbash}
    code~\cite{bro88}, while the ${\cal N}_{\rm max}=8$ energies were obtained
    using a serial version of the diagonalization code developed in
    Ref.~\onlinecite{gil12a}.}
\label{tabcom}
\end{table}
Next we investigate the quality of the PHFB wave function as a guiding wave function.
To this end, we compare in Table~\ref{tabcom} the CIMC energies using the
PHFB as the guiding wave function with the exact results for smaller values of
${\cal N}_{\rm max}$ and $N$, for which exact numerical diagonalization is possible~\cite{bro88,gil12a}.
We find that for an even (odd) number of particles the CIMC energies are within 1\% (2-3\%) of
the exact results. The accuracy of our method is comparable to that of the $\mathbf{r}$-space
fixed-node GFMC calculations for the same system~\cite{blu07}.

\begin{figure}[h!]
\includegraphics[width=\columnwidth]{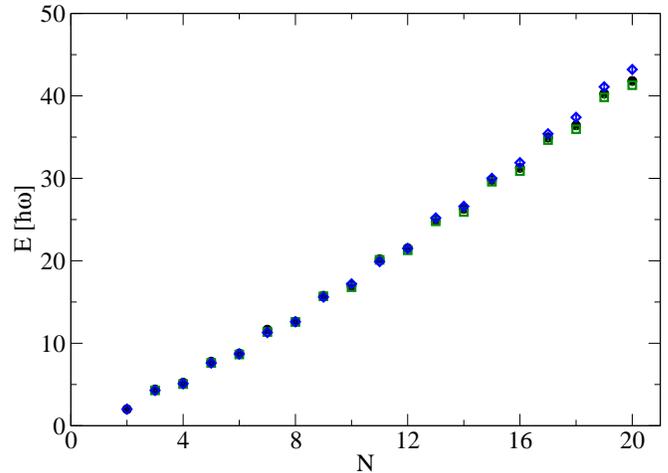}
\caption{The
    ground-state energy at unitarity for $N \leq 20$ particles. The CIMC results
    for ${\cal N}_{\rm max}=9$ (solid circles)  are compared with the
    results of Ref.~\onlinecite{blu07} (open squares) and Ref.~\onlinecite{cha07}
  (open diamonds). The statistical errors are smaller than the size of the symbols.}
\label{fignmax9}
\end{figure}
For larger systems exact numerical diagonalization is not possible. However,
$\mathbf{r}$-space fixed-node GFMC results are available~\cite{blu07,cha07}.
In Fig.~\ref{fignmax9} we compare the CIMC ground-state energies using $\mathcal{N}_{\rm max} = 9$
with these fixed-node GFMC calculations for $N \leq 20$. The CIMC ground-state energies using
${\cal N}_{\rm max} = 9$ oscillator shells are $0.5 - 3 \%$ higher than
those obtained in Ref.~\onlinecite{blu07}. The ground state energy estimates
 from Ref.~\onlinecite{cha07} are typically slightly higher than those in
Ref.~\onlinecite{blu07} but with much larger statistical error.
 At $\mathcal{N}_{\rm max} = 9$
the differences between the CIMC energies between successive values of
$\mathcal{N}_{\rm max}$ become smaller than their statistical errors. We have
not carried out any $\mathcal{N}_{\rm max} \to \infty$  extrapolations
in this work.

\begin{figure}[htbp]
 \includegraphics[width=\columnwidth]{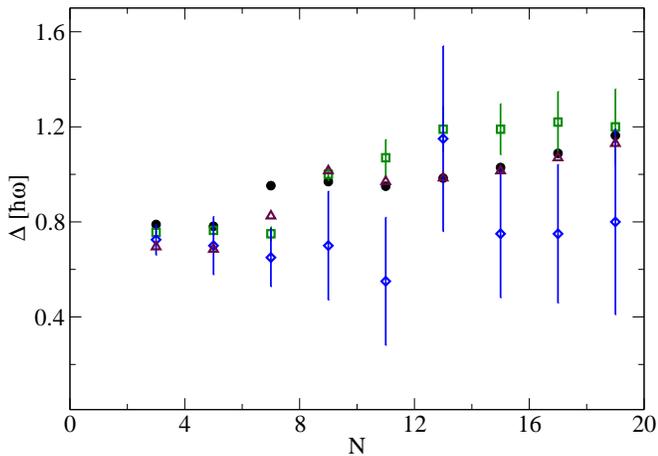}
  \caption{Energy-staggering pairing gaps versus number of atoms $N$. Our CIMC
    results for  $N_{\max} = 9$ (solid circles) are compared with similar gaps
    calculated in Ref.~\onlinecite{blu07} (open squares), Ref.~\onlinecite{cha07}
    (open diamonds) and Ref.~\onlinecite{bul07} (open triangles). The vertical bars describe the statistical errors.  The CIMC errors are smaller than the size of the symbols.}
  \label{figgaps9}
\end{figure}

In Fig.~\ref{figgaps9} we compare the energy-staggering pairing gaps $\Delta$
calculated in the CIMC method with the pairing gaps obtained from the two $\mathbf{r}$-space GFMC
calculations mentioned above~\cite{blu07,cha07}. We have also included the pairing gaps obtained
from the density functional theory calculations of Ref.~\onlinecite{bul07}.
The energy-staggering pairing gap for an odd-$N$ system is defined by
\be
\Delta = E_0(N) - \frac{1}{2}\left \{ E_0(N-1) + E_0(N+1) \right \}\;,
\ee
where $E_0(N)$ is the ground-state energy for $N$ particles.
Our results seem to be consistent with the results of Ref.~\onlinecite{blu07}
and Ref.~\onlinecite{bul07}, although the CIMC statistical error is much smaller than
the statistical errors of the $\mathbf{r}$-space GFMC calculations. The non-smooth behavior at $N=7$ is probably due to the shell closure at $N=8$.

For the results shown here we used $\sim 2 \times 10^5$ independent configurations for
the calculation of each energy. In most cases, the initial ensemble consisted of the non-interacting ground state. We also verified in a few test cases that the final result does not depend on the choice of the initial ensemble.
The calculations for ${\cal N}_{\rm max} \leq 4$
were carried out on a single processor and took up to a few hours. For ${\cal
N}_{\rm max} > 4$ the calculations were performed on a Linux cluster. The
largest calculation with ${\cal N}_{\rm max}=9$ and $N=20$ took about 900 cpu hours.
We choose $\delta \tau \sim 0.1$ for these calculations. With this choice the number of time steps required to project the ground state was $\sim 100$ and the number of decorrelation time steps was $\lesssim 5$. We did not find any significant dependence of the optimal $\delta \tau$,
projection time or decorrelation time on $\mathcal{N}_s$ or $N$.

The nominal scaling of the computational effort for the CIMC method with an AGP
guiding wave function is $\sim N^2 (\mathcal{N}_s-N)^2 \times N^3 \sim N^5 (\mathcal{N}_s-N)^2 $
where the factor $ N^2 (\mathcal{N}_s-N)^2$ comes from the maximal number of possible non-zero matrix elements in Eq.~(\ref{eqnpr1}) for a given $\bn$, while the factor $N^3$ is the computational
effort required to calculate a pfaffian or a determinant of an $N \times N$ matrix.
The pre-factor in this scaling can vary significantly depending on the type of interaction.
For example, in the system described above with interactions between up and down spins only, there are only
$(N/2)^2 \lbrace(\mathcal{N}_s-N)/2\rbrace^2$ non-zero terms in Eq.~(\ref{eqnpr1}) and we only have to calculate the determinant of an $N/2 \times N/2$ matrix. Thus, the computational effort in this case is about two orders of magnitude smaller than in the most general case.

It is interesting to compare our method with the constrained-path Monte Carlo~\cite{zha95} and the full configuration-interaction quantum Monte Carlo~\cite{boo09}
methods. Both of these methods perform a stochastic projection
of the ground-state wave function in Fock space similar to our method.
In the constrained-path Monte Carlo method, the Hamiltonian is `linearized' with the help of the Hubbard-Stratanovich transformation and the random walk is carried out in the continuous space of the auxiliary fields. The sign problem is circumvented with the help of a guiding wave function. However, the mixed energy estimate in this method is not an upper bound on the true ground-state energy.
In the full configuration-interaction quantum Monte Carlo method, the random walk is carried out in the discrete configuration space as in our
method. No guiding wave function is used in this method, and the
sign problem is mitigated using a configuration-annihilation algorithm.
However, the computational effort in this method scales with the size
of the many-particle space (i.e., it is exponential in $\mathcal{N}_s/N$),
although it is a fraction of the computational effort involved in a direct
diagonalization.

In conclusion, we have introduced a novel CIMC algorithm that
provides an upper bound for the ground-state energy of a finite fermionic system in the CI approach. We argue that the AGP class of wave functions provides a good choice for guiding wave functions.
We demonstrate CIMC for the trapped cold atom Fermi gas condensate at unitarity using the PHFB wave function as a guiding wave function.  For a
small number of particles and sufficiently small number of single-particle
orbitals, we find that the CIMC ground-state energies are within a few percent
of the exact results. Our CIMC results for number of
particles $N \leq 20$ are consistent with previous coordinate-space GFMC calculations.

We emphasize that the CIMC method described here is quite general and can be applied to other fermionic systems such as cold atoms in a deformed trap and for finite nuclei. It can also be used to calculate properties other than the ground-state energy, such as the average occupation numbers.
It will be interesting to determine whether the energy upper bounds can be improved by using other choices
for the geminals in the AGP wave function, or by using other classes of wave functions altogether.

\section*{Acknowledgements.}
We thank C.N. Gilbreth for useful discussions and for his help in providing some of the exact results in Table \ref{tabcom}.
This work was supported in part by the Department of Energy grant
DE-FG-0291-ER-40608. Computational cycles were provided by the facilities of the
Yale University Faculty of Arts and Sciences High Performance Computing Center.

\end{document}